\documentclass[11pt,twoside]{article}  
\usepackage{apn3conf}

\begin{document}   

%
%
%
%

\title{Effects of Stellar-Mass and the ISM on the Evolving Morphologies of
  Planetary Nebulae}  
\titlemark{Effects of Stellar Mass and the ISM on PN morphology}

%
%
%

\author{Eva Villaver}
\affil{Space Telescope Science Institute, 3700 San Martin Dr., Baltimore,
  21218 MD; {\it villaver@stsci.edu}}

\contact{Eva Villaver}
\email{villaver@stsci.edu}

%
%
%
%
%

\paindex{Villaver, E.}

%
%

\authormark{Villaver}


 
\begin{abstract}          

A significant step forward in the understanding of Planetary Nebula (PN) 
formation can be achieved by exploring the connection of  PN with stellar
evolution. 
In particular, the initial mass  of the star plays a crucial role,
as it determines the evolutionary timescales, the density structure of the
gas and the amount of energy injected into the nebula. Here we summarize our
study of the effects of stellar mass in PN formation. Our numerical
simulations include the evolution of the stellar wind for different initial
progenitor masses and the influence of the ISM. We also investigate how the
systemic velocity of the star with respect to its surrounding medium affects
the PN formation. We find that unless the star is moving, most of the mass
lost by PN  progenitors can be found in the low surface brightness extended
halos, where the stellar ejecta is mixed with ISM material. For a moving
central star, the interaction with the ISM considerably reduces the mass of
the  circumstellar envelope during the AGB and PN phases owing to ram
pressure stripping.
 
\end{abstract}

%
%
\section{Introduction}

Towards the end of the Asymptotic Giant Branch (AGB) phase, stellar evolution 
predicts episodic mass-loss increases, a consequence of the thermal pulses in
the star. Although the occurrence of this modulated mass-loss has played a
key role in 
the interpretation of the different shells  found around PNs and AGB stars,
hydrodynamic models describing the evolution of the wind in this phase still
need to be considered. Because of the inherent difficulties in the mass-loss
calculations and in recovering the history of mass-loss from observational
studies, the exact evolution of mass-loss during the AGB still remains
unknown. 
For this it is fundamental that the grid used in the numerical computations
allow the study of the whole stellar ejecta. It is only then, by comparing the
models with the PN structure  at large scales that constrains can be placed
on the treatment of mass-loss in models stellar evolution models. Previous
models have 
always truncated the computational grids to small scales.

Stellar evolution predicts that stars with main sequence masses
in the range of $\sim$1--5\,M$_\odot$ will produce PN, whilst PN
nuclei and white dwarfs mass distributions peak around 
0.6\,M$_\odot$. Since most of the mass-loss occurs on the AGB phase,
it should be easily observable as ionized mass during the PN stage. However,
observations of Galactic PNs reveal on average only 0.2\,M$_\odot$ of ionized
gas. 

Another important aspect that needs to be considered is the role played by
the stellar progenitor mass in the the PN formation.  The central star
provides the wind and the radiation field that determines the evolution of
the nebular gas during this phase. The details of the
post-AGB evolution of the star, and therefore the energy injected in
the nebular gas are mainly determined by its core
mass. Despite this, most of the numerical studies of PN evolution in the
literature have been restricted to a 0.6\,M$_\odot$ post-AGB evolutionary track. 

\section{The evolution of the stellar ejecta during the AGB phase.}

In Villaver, Garc\'{\i}a-Segura, \& Manchado (2002) we studied the
time-dependent hydrodynamics of the circumstellar  gas shells of AGB stars. We
directly used the results of stellar evolution (predictions for the wind
evolution from Vassiliadis \& Wood 1993) as inputs to the simulations. 

We find that the wind variations associated with the thermal pulses lead to
the 
formation of transient shells with an average lifetime of 20,000\,yr and,
consequently, do not remain recorded in the density or velocity structure of
the gas. The formation of shells that survive at the end of the AGB phase
occurs via two main processes: shocks between the shells formed by two
consecutive enhancements of the mass-loss or continuous accumulation of the
material ejected by the star in the interaction region with the ISM.  We do
not 
find the signature of discrete mass-loss recorded in the density or in the
velocity structure of the circumstellar envelope (CSE). Consequently, we
argued 
against the use of the different observed shells as dynamical clocks. 

The mass of the AGB stellar progenitors are usually estimated from observing
the CSEs, with the assumption that all the observed mass has been ejected by
the star. We find, however, that the final mass of the CSE contains a
significant fraction  of swept-up ISM material. Thus, in order to obtain
unbiased AGB-progenitor mass estimates, it is vital that this effect is taken
into account.  We also predict that the CSEs are a mix of the ISM and the wind
material that has been enriched by the stellar interior and brought to the
surface of the star by dredge-up processes. This effect should be taken into
account for abundance analysis.  According to our simulations based in the
mass-loss predictions of one particular set of stellar evolution models, large
CSEs (up to 2.5~pc) are expected around stars at the tip of the AGB.

\section{The connection between the stellar progenitor and the PN shell's
  evolution} 
 
To study PN formation we used a set of computational grids large  enough to
study the full stellar ejecta, and a set of grids small enough to resolve the
processes taking place close to the central star. In 
Villaver, Manchado, \& Garc\'{\i}a-Segura (2002) we considered  the evolution
of the post-AGB wind and the ionizing  radiation field for PN nuclei evolving
from progenitors with initial masses between 1 and 5\,M$_\odot$ (taken from
Vassiliadis \& Wood 1994). We show the importance of the dynamical effects of
ionization on the shell  evolution, which can account for the observed
disagreement between the kinematical ages and the age of the CSs for different
progenitor masses. We find that the evolution of the main shell is controlled
by the ionization front rather than by the thermal pressure provided by the
hot bubble during the early PN stages. 

The halos have sizes up to 2.3\,pc (more than twice the size of the main
shell) and are formed during the AGB phase. They produce H$\alpha$\
emissivities between 10 and 5000 times fainter than the main shell and
contain most of the total ionized mass lost by the stellar progenitors. Given
the low surface brightness of the halos, previous  observations have
underestimated the ionized masses in PNs. 

\section{The interaction of PNs with the ISM}
 
Several PNs show bow-shock structures, suggesting the interaction of the
nebular shell with the ISM while the star is moving. We have approached  the
problem of PN-ISM interaction through a realistic perspective by considering a
low-mass star  evolving during the AGB and PN phases while moving through the
ISM. In Villaver, Garc\'{\i}a-segura, \& Manchado (2003) we showed that even
the ejecta of a star with a systemic velocity of 20\,km\,s$^{-1}$ moving
through a low density medium will interact with it and form bow-shock
structures qualitatively similar to those observed.

An increase of ram pressure (due to e.g a higher velocity) 
leads to the development of instabilities that
lead to a partial fragmentation of the shell and to a more efficient mixing
with the 
ISM material. This fragmentation allows the UV radiation field to escape from
the nebula at certain locations. In Figure\,1 we show the logarithm of the gas
density during the latest stages of the  evolution for a star moving with
80\,km\,s$^{-1}$ through a low density ISM  (n$_{\rm o}$=0.05\,cm\,$^{-3}$). 
Figure\,1 shows only half of the $r-\theta$ plane, 
(where $r$ and $\theta$ are the radial and polar
coordinates respectively) the star is fixed in the grid and the ISM flows in
from the top to the bottom. We assume that the ISM moves relative to the star
perpendicular to the line-of-sight by fixing the position of the star at the
center of the grid and allowing the ISM to flow into the grid at the outer
boundary from 0$^{\circ}$ to 90$^{\circ}$. In Szentgyorgyi et al. (2003) we
considered a stellar motion with a velocity of 85\,km\,s$^{-1}$ in the study
of the PN NGC\,246.  We qualitatively reproduce the overall shape of the PN,
due to the interaction 
which we find consistent with what is expected in the fast interaction with a 
rarefied medium at the position of the PN in the Galaxy.

\begin{figure}
\plotone{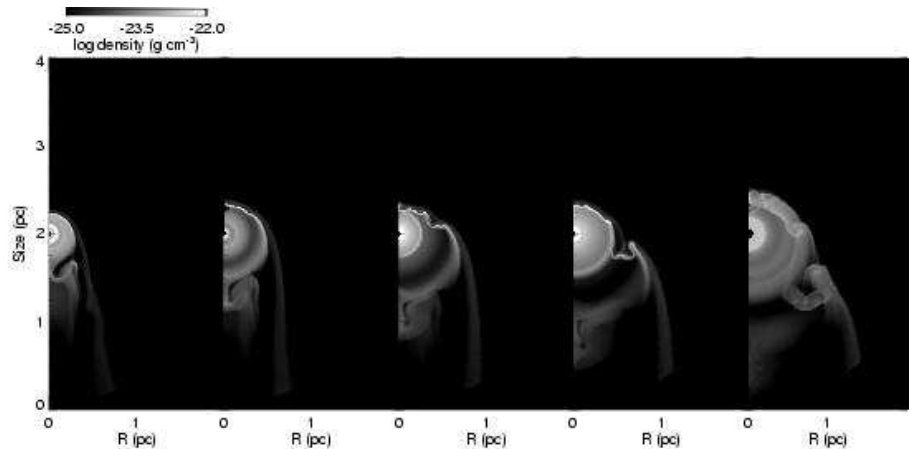}
\caption{Logarithmic density gas distribution around the
  central star during the AGB phase at times of 390,
  420, and 450$\times$10$^3$\,yr. The fourth panel correspond to the
  transition time and the last panel to a PN that is 8000\,yr old. 
  The relative movement takes place at a 
  velocity of 80\,km\,s$^{-1}$ with density 0.05\,cm\,$^{-3}$ and
  a temperature of 6000\,K.}  
\label{villaver1_f1}
\end{figure}

We find that due to ram-pressure stripping, most of the mass ejected during
the AGB phase is left downstream of the star in its motion, an effect that
might be able to account for the small amount of ionized mass recovered in PN
shells. We conclude that the interaction with the ISM plays a major role in
the PN formation process even during the early AGB evolution and 
it's interaction with the ISM cannot be studied using simple
ram pressure balance arguments.

\section{Summary}

We have studied the PN formation by following the evolution of the stellar
wind 
as predicted by stellar  evolutionary models and considering the influence of
the external ISM. We find that although the mass-loss history during the AGB
phase is very different for low- and high-mass progenitors, the final nebular
structure is very similar. The mass-loss during the AGB gives rise to the
formation of large shells (with sizes up to 3\,pc) that contain most of the
mass lost by the star plus an additional amount (up to 1\,M$_\odot$ of ISM
material) which is ISM material swept up by the stellar wind. We find that the
movement of the CS with respect to its surrounding medium considerably alters
the PN formation. The main effects of the interaction, apart from that on the
morphology, are  that the total size of the outer PN shell (halo) is reduced
considerably and most of the mass ejected during the AGB phase is stripped by
the ram pressure of the ISM and left in the downstream direction of the
stellar 
movement. The mass stripped away by the ISM when the star is moving might be
able by itself to account for the problem of the missing ionized mass in PNs.

%
%
%
%

\end{document}